\documentclass[twocolumn,showpacs,fleqn, pre]{revtex4-1}

	\usepackage{times}
	\usepackage{amsmath}
	\usepackage{gensymb}
	\usepackage{float}
	\usepackage{graphicx}
	\graphicspath{ {figures/} }
	\usepackage[colorlinks,hyperindex]{hyperref}
	\hypersetup
	{
		colorlinks=true,%
		citecolor=blue,%
		linkcolor=blue,%
		urlcolor=blue,%
	}
	\usepackage[all]{hypcap} 
   \DeclareMathOperator\erf{erf}  
     	
\makeindex

\begin{document}

\title{Using dislocations to probe surface reconstruction in thick freely suspended liquid crystalline films}

\author{J. A. Collett}
\email{jeffrey.a.collett@lawrence.edu}
\author{Daniel \surname{Martinez Zambrano}}
\affiliation{Department of Physics, Lawrence University, 711 E. Boldt Way, Appleton, Wisconsin, 54911}

\begin{abstract}
Surface interactions can cause freely suspended thin liquid crystalline films to form phases different from the bulk material, but it is not known what happens at the surface of thick films. Edge dislocations can be used as a marker for the boundary between the bulk center and the reconstructed surface.  We use noncontact mode atomic force microscopy to determine the depth of edge dislocations below the surface of freely suspended thick films of 4-n-heptyloxybenzylidene-4-n-heptylaniline (7O.7) in the crystalline B phase. 3.0 $\pm$ 0.1 nm high steps are found with a width that varies with temperature between 56$\degree$C and 59$\degree$C.  Using a strain model for the profile of liquid crystalline layers above an edge dislocation to estimate the depth of the dislocation, we find that the number of reconstructed surface layers increases from 4 to 50 layers as the temperature decreases from 59$\degree$C to 56$\degree$C. This trend tracks the behavior of the phase boundary in the thickness dependent phase diagram of freely suspended films of 7O.7, suggesting that the surface may be reconstructed into a smectic F region.
\end{abstract}


\maketitle

\section{Introduction}

The interactions at the free surface of liquid crystalline materials produce a number of surface ordering effects.  Free surfaces induce surface smectic ordering in the nematic phase \cite{PershanNA, PershanNA2}, surface hexatic ordering on smectic C surfaces \cite{Sirota3, Sorensen}, as well as surface crystallization on smectic A phases \cite{Pindak,Bishop1,Bishop}.  In all of these cases surface tension and surface orientational effects favor more ordered phases at the surface.  In this study, we present evidence that crystalline phases with modulated layers have smectic ordering for some depth below a free surface. 

\begin{figure}
\includegraphics[width=0.475\textwidth]{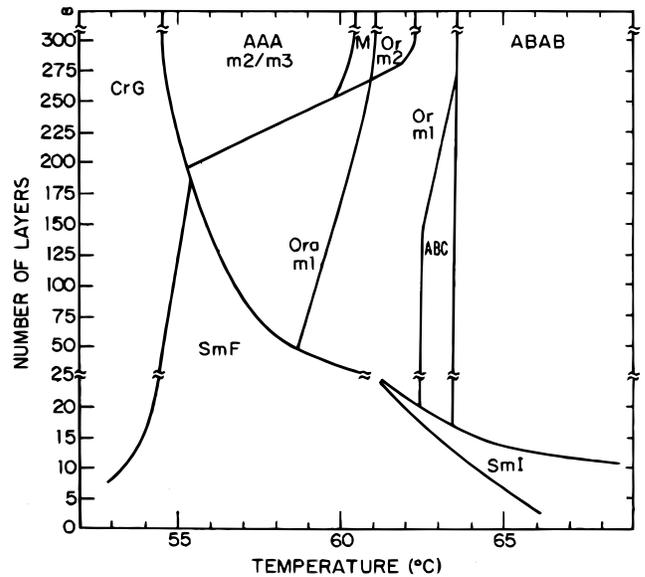}
\caption{Phase diagram of 7O.7 films as a function of temperature and thickness. In thin films crystalline phases are replaced by hexatic liquid crystalline structures.\label{fig:phasediagram}}
\end{figure}

The possibility of a smectic surface region in heptyloxybenzylidene heptylaniline (7O.7) is suggested by the thickness dependence of the phase sequence in freely suspended films.  Fig. \ref{fig:phasediagram}  shows the phase diagram of 7O.7 as a function of temperature and thickness \cite{Sirota2}. Bulk 7O.7 shows a crystalline B phase with ABAB stacking between 65$\degree$C and 69$\degree$C. As the temperature is lowered, the material forms orthorhombic-F, monoclinic-C, and hexagonal AAA structures as it cools. All of the phases below 63$\degree$C show modulated layer structures with the amplitude of the modulation increasing as the temperature decreases toward the transition to the crystalline G phase at 54.5$\degree$C. As the film thickness decreases these crystalline phases are eventually replaced by smectic I and smectic F phases, with the smectic F phase persisting in films over 100 layers thick between 55$\degree$C and 56$\degree$C. Optical birefringence suggests that there could be a smectic F layer on the surface of the crystalline phase. One model suggests that the smectic F phase is confined to the surface layer and that the entire film will only convert to the smectic F phase when the film is thin enough that there is too little crystalline phase to offset the free energy cost of a phase boundary \cite{Sirota}. The x-ray and optical techniques used did not have the ability to definitively determine the thickness of the smectic F region on top of the crystal. In this work we find that by examining edge dislocations near the surface that we can infer the depth of the smectic F phase.

We use a dynamic mode atomic force microscope to image the surface of thick ($\approx$100 $\mu$m) crystalline 7O.7 films at temperatures between 56$\degree$C and 59$\degree$C. In this temperature range the modulation of crystalline smectic layers would be measurable if there were no reconstruction at the surface. Height and width measurements of molecular steps (edge dislocations at or near the surface) determine the tilt of molecules at the steps and the distance of the dislocations from the surface while measurements of the topography away from the dislocation can determine whether the modulation is present at the surface. We find that the steps in the lower end of our temperature range are extraordinarily wide, suggesting that the dislocations are not at the surface. A strain model of defects in smectic phases can estimate the depth of the dislocation from the surface profile and finds that the defects are buried deeper under the surface as the temperature cools toward the transition to the crystalline G phase at 54.5$\degree$C.

\section{Experimental Details}

Surface topography was measured using a modified Park Systems XE-100 atomic force microscope (AFM) in dynamical imaging mode. We used sharp ($<$ 6 nm diameter) and super-sharp ($\leq$ 3 nm diameter) pyramidal silicon tips with a vibration amplitude of about 100 nm at 350 kHz. The system was calibrated with silicon carbide  samples, measuring step profiles with widths $<$ 5 nm and plateau surface roughness of 0.05 nm. The resolution is precise enough to detect layer modulations of the crystalline B phase that could have amplitudes up to 1 nm with a wavelength of about 10 nm.

Liquid crystal films were produced on a temperature controlled assembly mounted in the AFM. A support frame for producing the liquid crystal films was mounted on the heated stage of the microscope.  Films were drawn across a 6 mm diameter hole in a stainless steel plate with a beveled edge. The plate is attached to a heated aluminum assembly that has about a 2 mm gap between the film and the heated aluminum surface below. The film is heated conductively from the edge and radiatively from the aluminum assembly below. A LabView based PID temperature controller regulates the temperature of the plate with long term stability of 0.01$\degree$ C. Because our samples are sensitive to small temperature changes, we also developed a heated AFM tip assembly. We used two steel plates with a 0.5 mm insulating layer between them to support the AFM tips. One plate connects to the AFM head assembly. The other has the mounted cantilever, a heater, and a temperature sensor. The LabView controller regulates the tip temperature along with the sample temperature.   The film and tip sensors were calibrated in an ice-water bath at 0$\degree$C to minimize offsets between the film and tip temperatures.  When imaging, the tip and its mounting plate provide a radiatively coupled surface above the film to match the temperature of the surface below the film.  The controller ramps the temperature of the tip and film together whenever the temperature is changed. Thick liquid crystal films are drawn in the smectic A phase at 75$\degree$C and are cooled slowly into the crystalline B phases over a 12 hour period to produce high quality crystals. Film thickness is measured by focusing the optical microscope in the AFM first on visible terraces on the top surface and then on the bottom surface of the film. We use the position encoder on the microscope to measure the difference in positions. The films were 50 to 100 $\mu$m thick (15000 to 30000 molecular layers).
We collect data in the center of the film where it is most uniform, but we search for locations where we can find isolated steps to study.  Profiles of steps and regions around them are followed as a function of temperature.

\begin{figure}
\hspace*{-0.3cm}\includegraphics[width=0.5\textwidth]{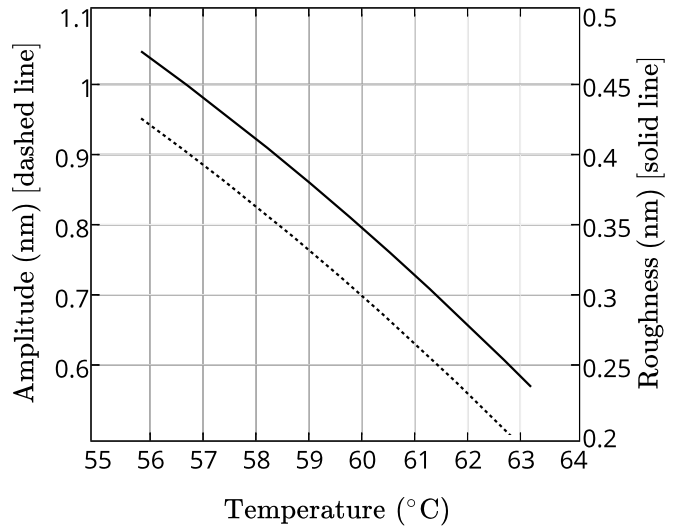}
\caption{Simulated amplitude(dashed line) and RMS roughness(solid line) of a modulated surface as a function of temperature.\label{fig:simulation}}
\end{figure}

\section{Results}

We present data demonstrating that between 56$\degree$C and 59$\degree$C the surface of freely suspended thick films ($>$ 15000 molecular layers) of 7O.7 shows no evidence of a modulated structure. In addition, molecular steps at the surface have been imaged and the height of the steps agrees with the upright molecular length of 7O.7. Finally, we examine how the width of the step profile varies with temperature.

We begin our investigation by developing quantitative expectations for an unreconstructed modulated surface. The phase diagram in Fig. \ref{fig:phasediagram} shows that between 55$\degree$C and 59$\degree$C, 7O.7 forms either a hexagonal AAA-m2 lattice structure, or a hexagonal AAA-m3. These phases have 2D modulations where the angle between the modulation wave vectors is 45$\degree$ for the AAA-m2 phase and 90$\degree$ for the AAA-m3 phase.  In these phases, x-ray data indicates that the amplitude of the modulation increases as the temperature decreases \cite{Sirota2}. The trend was identified by looking at the ratio of the intensity of the satellite peaks to the intensity of the main (001) Bragg peak. By simulating the x-ray structure factor for a single modulated layer, the amplitude of the modulation can be used to determine the peak intensity ratio.
Using the published intensity data, the amplitude of the modulation can be determined as a function of temperature. Assuming that the modulation is sinusoidal, we get the relationship shown in Fig. \ref{fig:simulation}. 

The amplitude grows as the temperature decreases until the sample reaches the transition to the crystalline-G phase at 54.5$\degree$C. While we show results for a sinusoidal modulation, we also simulated  triangle and sawtooth shapes. These shapes resulted in a higher amplitude for the modulation at the same intensity ratio, so we use the sinusoidal model to put a lower limit on the amplitude of the modulation. Since we anticipated that a static modulation may be difficult to image, we also calculated the RMS roughness that we'll expect from a modulated surface and found that the roughness prediction was nearly independent of the shape of the modulation.
At 56$\degree$C, we estimate that a sinusoidal modulation has an amplitude of 0.94 nm with a corresponding surface roughness of 0.47 nm. This is a significant fraction of the crystalline-B layer thickness of 3.05 nm.  

Our AFM measurements show no evidence of a modulation of this amplitude. The measured RMS surface roughness of the 7O.7 films is $0.03\pm0.04$ nm, well below what would be expected from an unreconstructed surface and right at the flatness limit of the instrument, suggesting that we have a flat reconstructed surface that differs from the interior of the crystalline film.

After confirming that there was no modulation on the surface, we imaged surface steps where the film thickness changes by one molecular layer.  If the surface is in the smectic F phase, the step height should be about 2.75 nm because of the molecular tilt. Instead, our collection of single steps from numerous locations on the surface of the film have heights of $3.0\pm 0.1$ nm, consistent with the molecular length of 7O.7(3.05 nm).

\begin{figure}
\hspace*{-0.3cm}\includegraphics[width=0.53\textwidth]{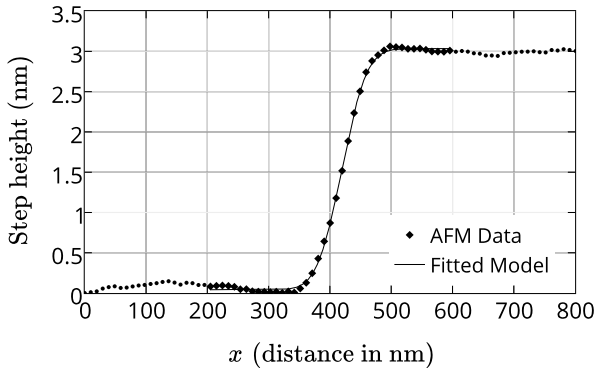}
\caption{Profile of a typical step on the film surface. The solid line represents a fit to the theoretical strain profile. \label{fig:singlestep}}
\end{figure}

 Fig. \ref{fig:singlestep} illustrates one of the step profiles taken at 56$\degree$C. Notice that the step width ($\sim$120 nm) is much greater than the molecular dimensions. This suggests that edge dislocations that produce the steps are not  on the surface, but instead are buried  a number of molecular layers under the surface.
\begin{figure}
\hspace*{-0.5cm}\includegraphics[width=0.57\textwidth]{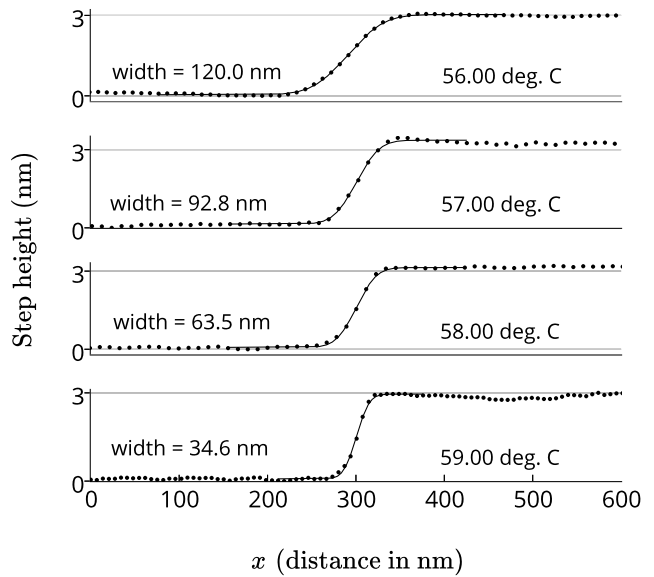}
\caption{Sample step profiles showing the change in step width with temperature. The dots show the AFM data and the solid line is the fit to an error function profile. The quoted widths are the range over which 95\% of the height change occurs.\label{fig:profiles}}
\end{figure}

Because the thickness dependence of the phase diagram in Fig. \ref{fig:phasediagram} changes with temperature, we explore how the step width varies as a function of temperature and find that the width decreases as the temperature increases. Fig. \ref{fig:profiles} shows several step profiles taken at different temperatures illustrating this trend.  If the edge dislocation were on the surface, the width should be comparable to the molecular size(3 nm). As a check on our resolution, we examined steps on a film of a crystalline B film of 4-n-butyloxybenzylidene-4-n-butylaniline(4O.8) and found step widths of about 16 nm, wider than the resolution limit but half the width of the sharpest 7O.7 steps.

One model that could explain these observations is that thick films of 7O.7 in the crystalline B phase have a surface region that is in a smectic F phase and that the edge dislocations that accompany changes in film thickness are located at the boundary between the crystalline and liquid crystalline regions. The model is developed in the following section.

\section{MODELING AND ANALYSIS}

Surface energy plays a crucial role in the dynamics of edge dislocations in smectic liquid crystals. In the absence of surface tension, an edge dislocation will slip to the surface to minimize the internal elastic energy \cite{Pershan_dislocation}.
On the other hand, when the surface energy is large compared with bulk elastic energy, the dislocation will be repelled from the surface \cite{Lejcek_1}.
In the case of a liquid crystal film on a flat substrate, surface interactions affect the position of the edge dislocation relative to the free upper surface of the film as well as the surface profile \cite{Holyst}. The effects of large surface tension are clearly present in AFM profiles of thin films of diblock copolymers deposited on a solid substrate \cite{Maaloum}.

\begin{figure}
\includegraphics[width=0.47\textwidth]{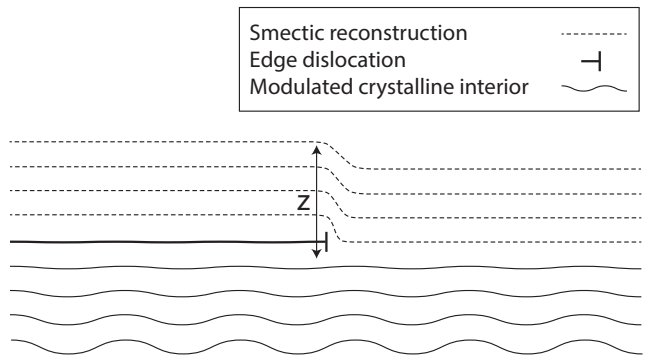}
\caption{Physical model of the dislocation at the interface between the crystalline and liquid crystalline regions. The flattening of the modulated layers is difficult to model, but this is a possible configuration. \label{fig:physicalModel}}
\end{figure}

We propose that our samples consist of a thick crystalline core with a liquid crystalline region of variable thickness on the surface. Dislocations near the surface of the sample are driven to the interface between the crystalline and liquid crystalline phases as shown in Fig. \ref{fig:physicalModel}. Since the height of the surface steps matches the extended length of the 7O.7 molecule, we assume that the dislocation is located on surface of the crystal and not in the liquid crystalline region because we would expect molecular tilts to reduce the layer spacing in that region. We show the crystalline layer modulations decreasing in amplitude as the interface is approached but this is purely speculative. To model the surface profile, we place an edge dislocation with Burger's vector $a$ at the interface and an image dislocation with an identical Burger's vector one layer below it. Here $a$ is the smectic layer spacing. This configuration will produce a strain solution with a flat layer at the plane of the interface \cite{Pershan_dislocation}.  Integrating the strain profile in the horizontal ($x$) direction at a constant height $z$ above the boundary gives the surface profile
\begin{equation}
    u(x,z) = \frac{a}{2} \erf\left(\frac{x}{2\sqrt{z\lambda}}\right).
\end{equation}
here $\lambda = \sqrt{K/B}$, where $K$ and $B$ are smectic elastic constants. If we assume that $\lambda \approx a$ as it is in other smectic phases, we can fit the profiles of the steps to infer the depth $z$ of the dislocation.

\begin{figure}[ht]
\hspace*{-0.36cm}\includegraphics[width=0.55\textwidth]{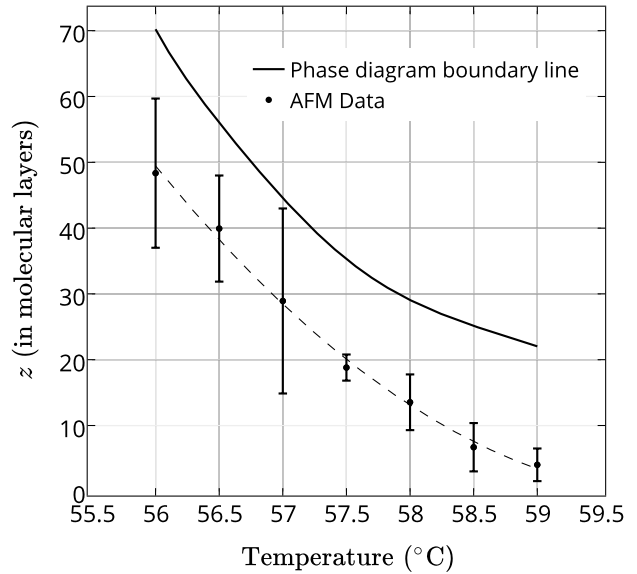}
\caption{Thickness of the smectic surface region as a function of temperature from fits to step profiles. The dashed line is a guide to the eye.  The solid line is derived by dividing the maximum smectic F film thickness at each temperature in Fig. \ref{fig:phasediagram} by 2. \label{fig:depth}}
\end{figure}

The model suggests that the thickness of the liquid crystalline surface layer increases as the temperature of the sample decreases.
Fig. \ref{fig:depth} shows the results of fits to many different step profiles taken at temperatures between 56$\degree$C and 59$\degree$C, with the thickness increasing from about 4  to nearly 50 smectic layers as the temperature of the sample  is reduced. Error bars represent one standard deviation for the set of measurements taken at each temperature. The simplest result that one might expect to find in this system is that there would be a smectic F surface region on the top and bottom of the film that is half the dimension of the thickest film that exists purely in the smectic F phase. The solid line in Fig. \ref{fig:depth} shows an estimate of that thickness.  We find surface regions thinner than that estimate but with thicknesses that follow the same trend with temperature.

\section{Discussion}

We have been able to show that there is a smectic surface region at the free surface of crystalline B 7O.7 samples with modulated layer structures.  Edge dislocations in the system localize at the boundary between the liquid crystalline and crystalline regions. Measurements of the surface profiles of steps provide a quantitative measure of the thickness of the reconstructed surface. Previous birefringence observations suggest \cite{Sirota2} that there is a smectic F surface layer above the modulated crystalline B phase; we can now measure its depth. A theory of modulated phases based on the energy of molecular tilt suggests modulations in the bulk may be destabilized by surface interactions in thin films \cite{Chen}. In that work the authors hypothesize that surface tension will lead to a flat boundary condition at the surface that will suppress the modulation, but they do not predict how the modulated crystalline B structure transforms into a surface smectic phase. While we can now measure the depth of the reconstruction, we still lack a detailed model of how the structure transforms at the interface.

We have identified several features of the surface smectic region. The thickness of the surface region increases with the the amplitude of the bulk modulation. This is not surprising since surface energy effects will have a bigger effect when the modulation is large. The depth of edge dislocations is a well defined function of temperature, suggesting that this represents an equilibrium structure. It takes several hours for the structure to equilibrate, but once it stabilizes, the measured step widths always converge to the same value at a given temperature. We see the same results when heating or cooling. Multiple films were prepared and all films gave the same results. The height of the surface steps in a stable sample always corresponds to the molecular length. Either the edge dislocation lies within the crystalline B structure or else the lowest section of the smectic region has no molecular tilt.  When new films are produced, we have seen steps with heights corresponding to the thickness of a tilted smectic F layer, but they eventually disappear.  This suggests that these dislocations either slip to the surface and move off the film or that they are repelled from the surface and move down to the boundary with the crystalline phase. Finally, we see that while surface energies are large enough to cause the reconstruction of the surface, they are not so large as to drastically broaden the profile of the step on the surface as is seen in the diblock copolymer system.

\section*{Acknowledgments}

The authors would like to acknowledge the contributions of Chris Hawley to the design of the free film stage and the heated AFM tip assembly.

\bibliography{References}

\end{document}